\documentclass{article}
\usepackage{dcase2022,amsmath,graphicx,url,times,booktabs, tabularx, xcolor}
\usepackage{siunitx}

\title{Foley Sound Synthesis at the DCASE 2023 Challenge}

\name{Keunwoo Choi$^{1}\sthanks{Equal contribution}$,
      Jaekwon Im$^{1, 2}{}^{*}
      $,
      Laurie M. Heller$^{3}{}^{*}$, 
      Brian McFee$^{4}$,
      }
\secondlinename{	  
      Keisuke Imoto$^{5}$, 
      Yuki Okamoto$^{6}$,
      Mathieu Lagrange$^{7}$,
      Shinnosuke Takamichi$^{8}$
      }
\address{$^1$ Gaudio Lab, Inc., Seoul, South Korea, \{keunwoo, jaekwon\}@gaudiolab.com\\          
        $^2$ KAIST, Daejeon, South Korea\\ 
        $^3$ Carnegie Mellon University, USA,
        laurieheller@cmu.edu\\
        $^4$ New York University, USA, brian.mcfee@nyu.edu\\
        $^5$ Doshisha University, Japan, keisuke.imoto@ieee.org\\
        $^6$ Ritsumeikan University, Japan, y-okamoto@ieee.org\\
        $^7$ CNRS, Ecole Centrale Nantes, Nantes Université, France, mathieu.lagrange@ls2n.fr\\
        $^8$ The University of Tokyo, Japan, shinnosuke\_takamichi@ipc.i.u-tokyo.ac.jp\\
 }

\begin{document}

\ninept
\maketitle

\begin{sloppy}

\begin{abstract}
The addition of Foley sound effects during post-production is a common technique used to enhance the perceived acoustic properties of multimedia content. Traditionally, Foley sound has been produced by human Foley artists, which involves manual recording and mixing of sound. However, recent advances in sound synthesis and generative models have generated interest in machine-assisted or automatic Foley synthesis techniques. To promote further research in this area, we have organized a challenge in DCASE 2023: Task 7 - Foley Sound Synthesis. Our challenge aims to provide a standardized evaluation framework that is both rigorous and efficient, allowing for the evaluation of different Foley synthesis systems. We received 17 submissions, and performed both objective and subjective evaluation to rank them according to three criteria: audio quality, fit-to-category, and diversity. Through this challenge, we hope to encourage active participation from the research community and advance the state-of-the-art in automatic Foley synthesis. In this paper, we provide a detailed overview of the Foley sound synthesis challenge, including task definition, dataset, baseline, evaluation scheme and criteria, challenge result, and discussion.
\end{abstract}

\begin{keywords}
Generative models, DCASE, sound synthesis
\end{keywords}

\section{Introduction}
\label{sec:intro}
Recent years have seen remarkable progress in generative models, with applications in a variety of fields including image generation~\cite{ramesh2021zeroshot}, text generation~\cite{openai2023gpt4}, music generation~\cite{pasini2022musika, agostinelli2023musiclm, donahue2018adversarial}, and sound generation~\cite{borsos2022audiolm, liu2023audioldm}. Models like these are capable of generating high-quality and diverse samples, and have been widely adopted in both academia and industry. In particular, sound generation has gained increased attention in recent years, with advances in sound synthesis and generative models enabling the creation of realistic and diverse audio content.

Sound synthesis plays a crucial role in enhancing the auditory perception of multimedia content, such as movies, music, and videos. Automatic or machine-assisted Foley synthesis has the potential to greatly streamline the process of creating these sound effects, freeing up time and resources for multimedia content creators.

To encourage further research and development in the field of automatic Foley synthesis, we developed a challenge that aims to provide a standardized evaluation framework for different systems. Challenges have been shown to be an effective way to motivate the development of machine learning models, particularly in the early stages of a research area. We believe that this Foley sound synthesis challenge can play a critical role in advancing the state-of-the-art in automatic Foley synthesis. This challenge was held as part of the international Detection and Classification of Acoustic Scenes and Events 2023 Workshop. The topics discussed in this introduction are also covered in a proposal document~\cite{choi2022proposal}.

\section{Problem and Task Definition}
We defined the problem of this challenge as `category-to-sound' generation. The category is chosen in one of the selected seven categories - \emph{dog bark}, \emph{footstep}, \emph{gunshot}, \emph{keyboard}, \emph{moving motor vehicle}, \emph{rain}, and \emph{sneeze/cough}. The sound is specified as a 4-second mono audio snippet with a sampling rate of 22,050~Hz. 

As this was the first year of this challenge, we chose the input of the system to be a sound category rather than text input with natural language. This simplification was made to ease the organizing effort such as defining the problem and the evaluation scheme, collection of dataset, etc. We also intended this to lower the bar for participation, especially from academia, as category-based systems would require less data and computational resources than free text inputs. Similarly, limiting the problem to the seven categories clarified the subjective evaluation criteria. The seven categories were chosen so that i) the categories are useful for media creation, ii) it is feasible to collect a reasonable quantity of training/evaluation sounds with manual review, and iii) the generated sounds are easy to assess for the evaluators. 

Despite this simplification, our intention for this challenge is to build towards generalizable and potentially useful approaches in the real world. In this regard, we specified the submitted systems should not simply copy-paste an existing sound, i.e., the systems should be generative, not retrieving. 

Our goal is to motivate the development of new methods for Foley synthesis. Because the volume of data can be instrumental in qualitative improvements across many areas of ML, we created two challenge tracks: one in which participants are free to augment their training data with external sources (Track A), and the other in which only the provided development dataset is allowed (Track B).
To enhance the efficiency of the challenge, we also provided two pre-trained models, HiFi-GAN~\cite{hifigan} and VQ-VAE~\cite{vqvae}, for Track B. These models were trained using the official dataset.

For a fair and correct evaluation, we required the participants to submit their model embedded in a Google Colab notebook template\footnote{\url{https://colab.research.google.com}}. This provided an easy, familiar, and verifiable way for participants to share models while resolving any dependency issue for the organizers, at least within the time frame of the challenge. 

\section{Official Dataset and Baseline}

The development dataset used in this task consists of 6.1 hours of audio excerpts, each annotated with one of seven distinct sound classes: footstep, sneeze/cough, rain, dog bark, moving motor vehicle, gun shot, and keyboard. We selected the categories by considering an urban sound taxonomy~\cite{urbansound8k}. The seven sound categories were selected evenly from each top-level group (`human', `nature', `mechanical'), except for `music.' There is no overlap in the low-level groups between the sound categories. 

We collected the data from UrbanSound8K~\cite{urbansound8k}, FSD50K~\cite{fsd50k}, and BBC Sound Effects.\footnote{\url{https://sound-effects.bbcrewind.co.uk/}} To select the appropriate audio clips for our challenge, we followed a two-step process. First, we gathered audio samples that were annotated with labels closely related to one of the seven sound categories. Second, to ensure consistency in the challenge, we pre-processed the audio to mono 16-bit 22,050 Hz and either zero-padded or segmented it to a length of 4 seconds, a duration found sufficient for human recognition of class and audio quality. This pre-processing step was applied before selection, as the audio events comprise only a small portion of the total audio length.

To ensure the quality of the dataset, we carefully selected the audio clips for each category based on their relevance, variety, and clarity. One organizer manually selected the collection of excerpts, each of which was verified by a different organizer to ensure accuracy and clarity. Overall, we selected 5,550 labeled sound excerpts, with the number of sounds per category ranging from 681 to 900.

We divided the dataset into a development dataset and an evaluation dataset. Although the number of audio samples varies across sound classes, we ensured that the evaluation set had a consistent number of 100 audio samples per category. This decision was made to ensure that the evaluation set had a diverse range of sounds and was not too small. We also made sure that the partitions were stratified, so no source recording provided clips in both the development and evaluation sets, even if there were multiple excerpts from the same longer recording.

As a baseline system, we implemented a  model~\cite{baseline} composed of three independently trained modules: PixelSNAIL~\cite{pixelsnail},
VQ-VAE~\cite{vqvae}, and HiFi-GAN~\cite{hifigan}.
The first module, PixelSNAIL, is an autoregressive model that maps a sound category input to a time-frequency representation.
The second module, VQ-VAE, transforms the PixelSNAIL output into a Mel spectrogram through a compressed, latent vector encoding.
The final module, HiFi-GAN, transforms the VQ-VAE output (Mel spectrogram) into a time-domain digital audio signal.

We selected the model as our baseline system for the following reasons. First, the modules were assigned the reconstruction task and the generation task separately, enhancing the whole architecture's explainability. Second, the participants were allowed to reuse some of the modules. Since each module was trained independently, improving the performance of the system can be achieved by modifying the structure or scheme of specific modules while keeping the remaining modules unchanged.

\section{Evaluation}
\label{subsec:eval_plan}
Even for objective tasks such as classification and detection tasks, it is challenging to provide unambiguous annotations and unbiased evaluation metrics. Multiple evaluation metrics may be necessary, but it can complicate the ranking of participants. \cite{mesaros2016metrics}.
With generative tasks such as the one considered in this challenge, the problem is even more difficult, as the produced data is not a set of labels, but audio, whose qualities must be assessed. This matter is far from being solved and is currently undergoing active research \cite{vadillo2022human}. Recognizing this as a challenge, we opted for a pragmatic combination of objective and subjective evaluation protocols as proposed in~\cite{choi2022proposal}.

In detail, we chose a two-step procedure. The first step considers objective metrics to get a first ranking of the proposed systems. Due to the constraints on human listening time for subjective ratings, in each track, only the top four entries were then considered for the second step with a subjective evaluation.

We decided to measure the following qualities:
\begin{enumerate}
    \item \textbf{Perceptual Audio Quality}: The degree of clarity of sound, free from any artifacts, fuzziness, degradation, distortion, and noise. 
    \item \textbf{Fit-to-category}: The degree to which a sound is recognized as belonging in the intended category. 
    \item \textbf{Diversity}: The degree to which a system is able to produce a diverse set of sounds.
    
\end{enumerate}

Evaluation of the above qualities typically involves high-level perceptual and cognitive processing by humans and thus cannot be evaluated by simple computational means. For this reason, we chose to complement the objective evaluation with subjective metrics. Although essential, subjective evaluation comes with some constraints. Humans can give different ratings depending upon the context of a sound they hear, and can experience fatigue. For the latter reason, only a subset of audio samples can be presented for subjective rating. To make the sure the context is similar across raters (and potentially, across future contests), the audio samples should include some ``anchors," i.e. sounds which clearly have a very low and/or high quality; anchors help to psychologically anchor the ratings and also serve as a check on the quality of the rater~\cite{itu20031534}.

\subsection{Step 1: Objective Evaluation}

We adopted Fréchet Audio Distance (FAD)~\cite{fad}, a reference-free, lower-the-better, evaluation metric. FAD calculations were performed for each category. Systems were then ranked based on the average FAD across seven categories, and only the 4 top-performing systems per track were considered for the second step, due to time limitations of the subjective evaluation.

\subsection{Step 2 : Subjective Evaluation}

The subjective evaluation was operated in two steps. The first was an online survey that measured the fit-to-category and perceptual audio quality. The fit-to-category asked the listener to use their general notion of the sound category and was not restricted to referencing the exact sounds in the development set, nor was it based on the number of sound events in a file. These tasks were performed on 20 sounds from each category, along with a set of anchors taken from the development set and baseline system. The second step was a measure of category diversity.

The selection of the 20 representative sounds was done as follows. OpenL3 embeddings of all the samples were computed and a k-means clustering with $k=20$ was conducted on them \cite{cramer2019look}. The 20 ``medoid" representative sounds are selected as the ones with the smallest Euclidean distance to the centroid in the embedding space.

After listening to each sound, the rater was asked to rate two scales to indicate both its perceptual audio quality and its fit-to-category, as defined in Section~\ref{subsec:eval_plan}.
For both scales, raters selected among 11 levels, with 0 being an unusable sound and 10 being the top of an absolute scale (the best possible, as opposed to the best of this contest). Re-listening to the sound was permitted. This procedure was more appropriate for category fit judgments than MUSHRA \cite{itu20031534} because each sound was unique and different sounds could fit a category equally well.

Before rating a category, the rater listened to 6 representative sounds of the category from the development set. The high and low quality/fit anchor sounds, respectively, were hand-picked from the evaluation set and our baseline system. These sounds were not identified as anchors in the survey and were embedded in the main test at random locations. Entries from Track A and B were intermixed so that their relative quality would be apparent, even though the competition rankings are separated within each track. The order of trials was counterbalanced across test conditions.

\subsection{Execution}
All of the challenge participants performed the ratings on perceptual audio quality and fit-to-category for 4-7 categories, for a total duration of about 3-6 hours. After each category, the listener could take a break. 

All participants listened to the same sounds. Thus, participants who submitted one of the finalist systems actually rated sounds from their own systems but their self-ratings were removed by the organizers before computing results. This allowed us to streamline the rating system while removing potential rating bias. 

Rating at least 4 categories was required to be eligible for a prize. This requirement ensured that we had a fair distribution of teams doing ratings and enough ratings per sound. Additionally, some organizers rated sounds. This combined effort resulted in 10-15 independent ratings per sound. 93 separate category ratings were completed which took approximately 47 hours. Two of the 93 ratings were omitted at the start of the data analysis because they mis-rated 5 or more of the 12 quality-check trials (in both cases, giving a rating of low quality \& good fit to an anchor sound that had a high quality \& poor fit, indicating that they had confused the two scales).  The anchors that had low quality tended to get a poor fit rating, so we did not use those as an exclusion criterion. Appropriate ratings were given for anchors that had high quality \& low fit, high quality \& good fit, and low quality \& poor fit (4 of each type).

To validate the protocol as well as the software stack, a pilot study was carried out with the outputs of the baseline system in which the listeners were the organizers. During the evaluation phase, the test was advertised to relevant mailing lists. In this version, only one 30-minute category rating task was proposed to the listener, using a scheme to distribute the ratings across categories. 

Finally, as our second step, we also performed a subjective test on Diversity. Diversity is a ``set-based" quality, meaning that a set of generated audio files are mandatory for measuring it. For this reason, Diversity could not be evaluated within the above discussed listening test, whose stimuli are considered independently. For each system and each category, an organizer who did not participate in the ratings generated a continuous audio file sequencing the 20 representative sounds per system. Each file was given a name specifying the category and an obfuscated version of the system id. The diversity rating task took about 1.5 hours. Four other organizers, blind as to which systems they were rating, rated the diversity of the sounds per file from 0 (All the sounds appear to be identical) to 10 (Extremely large range of sounds).

Considering that 1) diversity may be less important than quality and fitness and 2) this quality has been not as rigorously tested in this edition of the challenge as the two other qualities, organizers decided in advance that the diversity ratings were weighted half as much as each of the audio quality and category fit ratings.

\section{Results}

We provided a Colab notebook as a starting point to implement submissions. We  received 42 systems in total, including 11 systems in Track A \cite{ChonGLI2023, GuanHEU2023, Leemaum2023, ScheiblerLINE2023, YiSURREY2023} and 31 systems in Track B \cite{BaiJLESS2023, ChangHYU2023, ChunChosun2023, ChungKAIST2023, JungKT2023, KamathNUS2023, LeeMARG2023, Leemaum2023, PillayCMU2023, QianbinBIT2023, QianXuBIT2023, WendnerJKU2023, XieSJTU2023}. 
We removed disqualified submissions that failed to run on standard Colab instances in a reasonable period of time. Before disqualification, we had a 4-day review period permitted for trivial bug fixes but did not allow changes in parameterization of the submitted systems.  

The six selected sounds from each category of the development set, which served as initial category referents, received an average perceptual quality rating of 8.6 and an average fit of 9.0. (Quality / fit averages for categories 1 through 7 were: 8.3 / 8.7, 8.2 / 8.5, 8.9 / 9.4, 9.12 / 9.3, 8.7 / 9.1, 8.0 / 8.6, 9.1 / 9.5). Diversity was not rated. These perceptual ratings validate our task because our listeners gave high ratings to genuine recorded sounds, but they do not represent the average quality of the development set, because these referent sounds were preselected to have good quality and diversity based on organizer judgments.

With the remaining 36 working systems submitted by 17 teams, we generated 700 audio samples from 9 and 27 systems for tracks A and B, respectively. The audio samples are available online\footnote{\url{https://zenodo.org/record/8091972}}. 

As all the scores (FAD scores per category, subjective test results on audio quality, fit-to-class, and diversity) were released on the DCASE official website,\footnote{\url{https://dcase.community/challenge2023/task-foley-sound-synthesis-results}} we analyze the evaluation results in this section. 

In Fig. \ref{fig:scatter1}, the FAD scores of 17 systems are plotted. The $(x, y)$ position represents the average FAD score computed on the development set (FAD-Dev) and the evaluation set (FAD-Eval), respectively. The width and height of each rectangle represents the (scaled) standard deviation over 7 categories for both sets, respectively.

First, most of the systems show better (lower) FAD-Dev than FAD-Eval, with the exception of \cite{ChonGLI2023}. This is expected, as the training would be at least partially based on the development set. Second, it turns out that FAD-Dev is a noisy measure to predict FAD-Eval. This is not surprising as the final objective measure (FAD-Eval) contained new sounds to prevent overfitting. Third, comparing the top systems of track A and B, several systems in track B showed better performance on FAD-Dev, but not in FAD-Eval. This shows the difficulty of training a system with the limited amount of data permitted in track B. 

In Fig. \ref{fig:scatter2}, the top 8 systems and the baseline system are plotted by their final ranking determined by a listening test as well as FAD-Eval and FAD-Dev. On the left, the scatterplot shows the importance of subjective tests. The Spearman's rank correlation coefficient of the ranking by FAD-Eval and the final ranking is only `0.238'. On the right, with FAD-Dev, the coefficient is somewhat higher, `0.524'.  

We established that subjective perceptual sound qualities were not entirely predicted by objective FAD scores.  In addition, we established that the three perceptual metrics were interrelated, but each had a unique contribution. Within each category, the correlations between average rating scores of finalist systems of audio quality and category fit were very strong (average across all categories was $r=0.98$); however, when quality \& fit ratings from individual trials were correlated within each category, the average correlation was less extreme ($r=.75$), showing that raters were not giving identical answers to both questions. Our anchor trials showed that the raters did know how to distinguish the two qualities, because they appropriately rated the category-inappropriate sounds with good audio quality. On the other hand, we also found that raters gave all-around low ratings to the category-appropriate sounds with poor audio quality.  Because sound recognition was essential for judging category fit,  it is plausible that good audio quality was required before being able to give a high category fit rating. Average diversity (within each category,  across finalist systems) had a strong relationship to category fit ($r=0.70$); nonetheless, half of the variance in diversity ratings was independent of quality/fit.

The perceptual ratings of the quality/fit of all the systems were plausible, with the highest average ratings obtained for the sounds from the development set, and the lowest for our baseline system. The submitted systems had intermediate ratings, showing that there is room for improvement in this challenge.

To summarize, there are expected mismatches between the objective evaluation for the provided sounds (FAD-Dev) and the sounds held back by the organizers (FAD-Eval);  importantly, objective evaluation metrics did not completely align with subjective evaluation (final ranking). This justifies two of our choices for the evaluation scheme: i) receive submissions in the form of a system (code) instead of sounds, and ii) run a subjective evaluation.

\begin{figure}[t!]
  \centering
  \includegraphics[width=0.65\columnwidth]{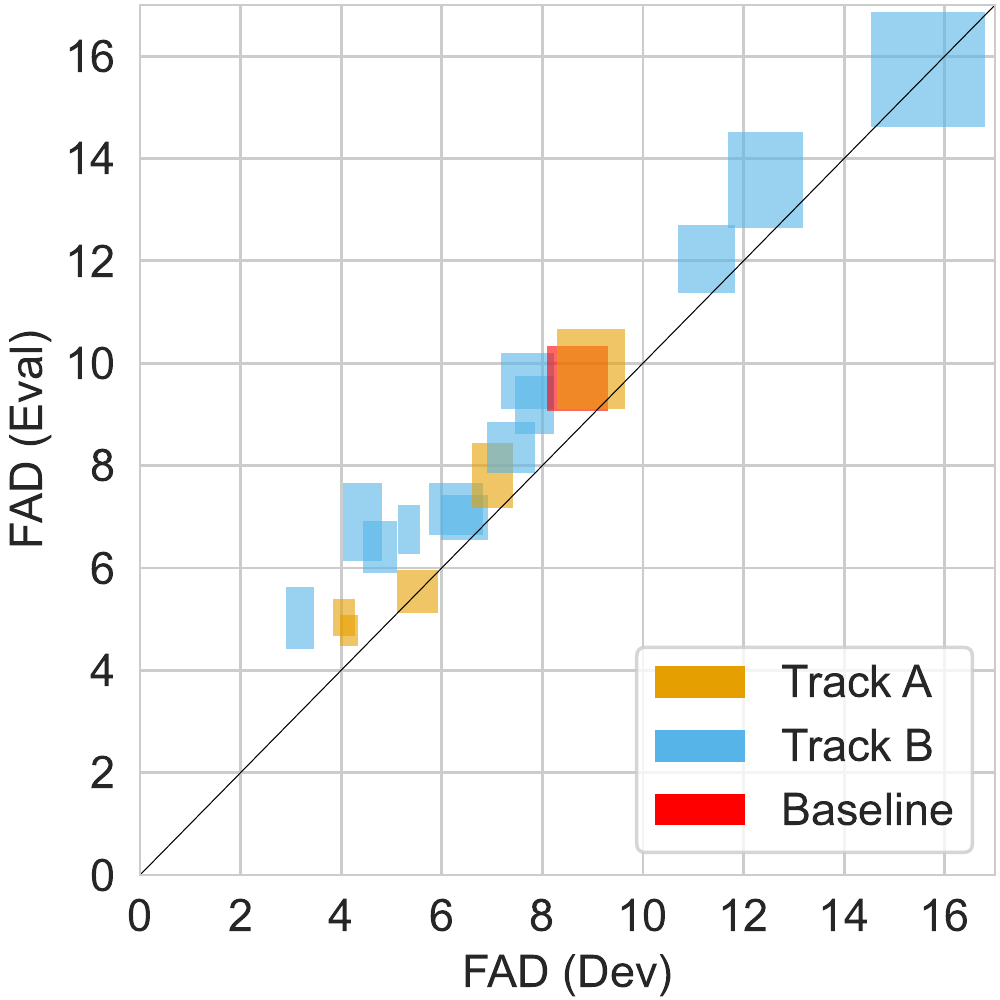}
  \caption{FAD Scores on the development set vs the evaluation set, computed on the 17 submitted systems and the baseline system.}
  \label{fig:scatter1}
\end{figure}

\begin{figure}[t]
  \centering
  \includegraphics[width=\columnwidth]{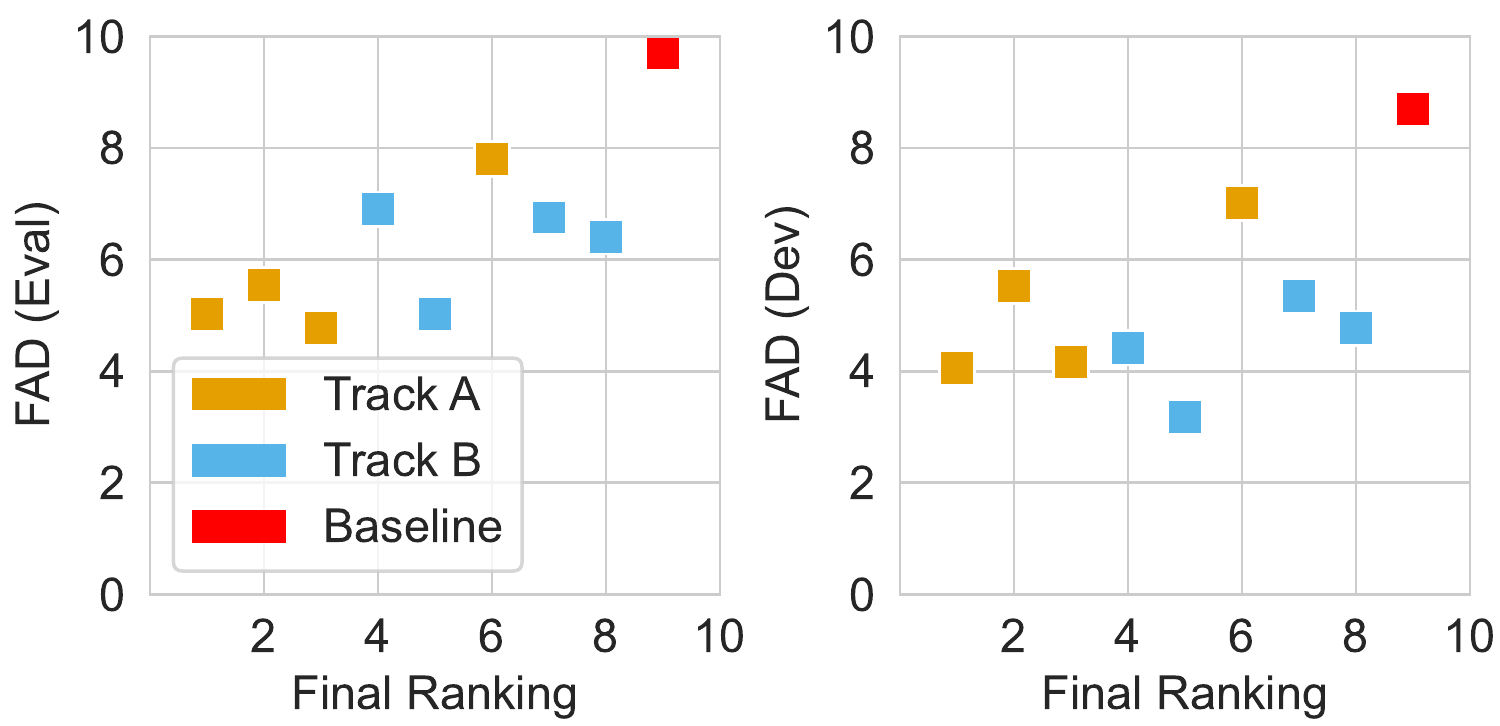}
  \caption{FAD scores on the development set and the evaluation set vs. the final ranking determined by the listening tests.}
  \label{fig:scatter2}
\end{figure}

\begin{figure}[t!]
  \centering
  \includegraphics[width=\columnwidth]{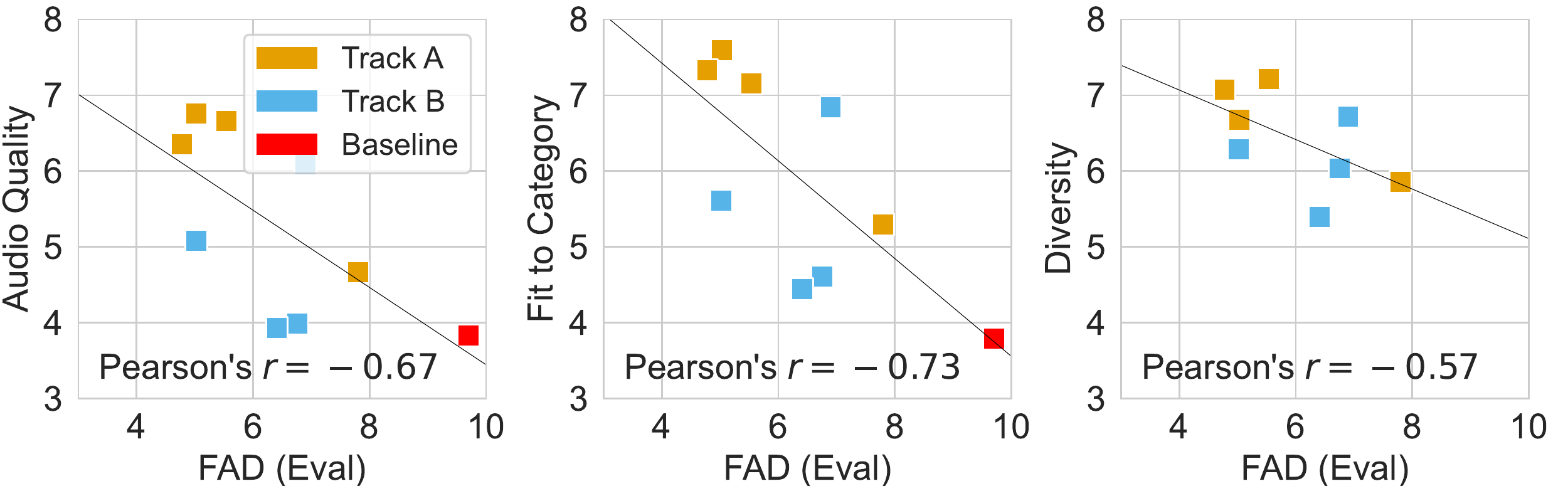}
  \caption{Relationships between objective measure (FAD-Eval) and subjective tests.}
  \label{fig:scatter3}
\end{figure}

\section{Conclusion} In this paper, we presented a challenge for automatic Foley sound synthesis aimed at promoting further research and development in generative AI for sound. We have provided a detailed overview of the challenge, including task definition, dataset requirements, evaluation criteria, a baseline implementation, and analysis of the results. Through this challenge, we believe we have achieved our goal --- to encourage active participation from the research community and advance the state-of-the-art in automatic Foley synthesis. Although it was the first year of the challenge, we received substantial submissions in both of the tracks. We also performed the generation and evaluation of the submitted systems successfully. 

In both tracks, the best performing systems were based on deep learning, with a sequence of a diffusion model for spectrogram generation and HIFI-Gan \cite{hifigan} for phase reconstruction.

There have been difficulties as well. Our analysis showed the necessity of performing a subjective evaluation and running inference by ourselves. Unfortunately, both are costly; in total, about 47 hours were spent for the evaluation of 8 systems and about 471 A100 GPU hours for the inference. With permission, we released all the generated sounds as well as their subjective/objective scores on Zenodo, hoping to enable more analysis and even subjective quality prediction models based on the data.

In the future, we hope that the standardized evaluation framework provided by this challenge will help to facilitate comparisons between different Foley synthesis systems. It is already apparent that more sophisticated Foley sound synthesis will be possible in the near future with text-input, video-input, etc. We hope our challenge will ultimately lead to the development of more effective and efficient techniques.

\section{Acknowledgements}
Computing resources for sound generation were covered by Gaudio Lab, Inc. We thank all the raters who did the subjective evaluation.

\bibliographystyle{IEEEtran}
\bibliography{refs}

\begin{thebibliography}{10}
\providecommand{\url}[1]{#1}
\def\UrlFont{\rmfamily}
\providecommand{\newblock}{\relax}
\providecommand{\bibinfo}[2]{#2}
\providecommand\BIBentrySTDinterwordspacing{\spaceskip=0pt\relax}
\providecommand\BIBentryALTinterwordstretchfactor{4}
\providecommand\BIBentryALTinterwordspacing{\spaceskip=\fontdimen2\font plus
\BIBentryALTinterwordstretchfactor\fontdimen3\font minus
  \fontdimen4\font\relax}
\providecommand\BIBforeignlanguage[2]{{%
\expandafter\ifx\csname l@#1\endcsname\relax
\typeout{** WARNING: IEEEtran.bst: No hyphenation pattern has been}%
\typeout{** loaded for the language `#1'. Using the pattern for}%
\typeout{** the default language instead.}%
\else
\language=\csname l@#1\endcsname
\fi
#2}}

\bibitem{ramesh2021zeroshot}
A.~Ramesh, M.~Pavlov, G.~Goh, S.~Gray, C.~Voss, A.~Radford, M.~Chen, and
  I.~Sutskever, ``Zero-shot text-to-image generation,'' in \emph{International
  Conference on Machine Learning}.\hskip 1em plus 0.5em minus 0.4em\relax PMLR,
  2021, pp. 8821--8831.

\bibitem{openai2023gpt4}
OpenAI, ``{GPT-4} technical report,'' \emph{arXiv preprint arXiv:2303.08774},
  2023.

\bibitem{pasini2022musika}
M.~Pasini and J.~Schl{\"u}ter, ``Musika! fast infinite waveform music
  generation,'' in \emph{ISMIR 2022 Hybrid Conference}, 2022.

\bibitem{agostinelli2023musiclm}
A.~Agostinelli, T.~I. Denk, Z.~Borsos, J.~Engel, M.~Verzetti, A.~Caillon,
  Q.~Huang, A.~Jansen, A.~Roberts, M.~Tagliasacchi, \emph{et~al.}, ``{MusicLM}:
  Generating music from text,'' \emph{arXiv preprint arXiv:2301.11325}, 2023.

\bibitem{donahue2018adversarial}
C.~Donahue, J.~McAuley, and M.~Puckette, ``Adversarial audio synthesis,'' in
  \emph{International Conference on Learning Representations (ICLR)}, 2019.

\bibitem{borsos2022audiolm}
Z.~Borsos, R.~Marinier, D.~Vincent, E.~Kharitonov, O.~Pietquin, M.~Sharifi,
  D.~Roblek, O.~Teboul, D.~Grangier, M.~Tagliasacchi, \emph{et~al.}, ``Audiolm:
  a language modeling approach to audio generation,'' \emph{IEEE/ACM
  Transactions on Audio, Speech, and Language Processing}, 2023.

\bibitem{liu2023audioldm}
H.~Liu, Z.~Chen, Y.~Yuan, X.~Mei, X.~Liu, D.~Mandic, W.~Wang, and M.~D.
  Plumbley, ``{AudioLDM}: Text-to-audio generation with latent diffusion
  models,'' \emph{arXiv preprint arXiv:2301.12503}, 2023.

\bibitem{choi2022proposal}
K.~Choi, S.~Oh, M.~Kang, and B.~McFee, ``A proposal for foley sound synthesis
  challenge,'' \emph{arXiv preprint arXiv:2207.10760}, 2022.

\bibitem{hifigan}
J.~Kong, J.~Kim, and J.~Bae, ``{HiFi-GAN}: Generative adversarial networks for
  efficient and high fidelity speech synthesis,'' \emph{Advances in Neural
  Information Processing Systems}, vol.~33, pp. 17\,022--17\,033, 2020.

\bibitem{vqvae}
A.~Van Den~Oord, O.~Vinyals, \emph{et~al.}, ``Neural discrete representation
  learning,'' \emph{Advances in neural information processing systems},
  vol.~30, 2017.

\bibitem{urbansound8k}
J.~Salamon, C.~Jacoby, and J.~P. Bello, ``A dataset and taxonomy for urban
  sound research,'' in \emph{Proceedings of the 22nd ACM international
  conference on Multimedia}, 2014.

\bibitem{fsd50k}
E.~Fonseca, X.~Favory, J.~Pons, F.~Font, and X.~Serra, ``Fsd50k: an open
  dataset of human-labeled sound events,'' \emph{IEEE/ACM Transactions on
  Audio, Speech, and Language Processing}, vol.~30, pp. 829--852, 2021.

\bibitem{baseline}
X.~Liu, T.~Iqbal, J.~Zhao, Q.~Huang, M.~D. Plumbley, and W.~Wang, ``Conditional
  sound generation using neural discrete time-frequency representation
  learning,'' in \emph{2021 IEEE 31st International Workshop on Machine
  Learning for Signal Processing (MLSP)}.\hskip 1em plus 0.5em minus
  0.4em\relax IEEE, 2021, pp. 1--6.

\bibitem{pixelsnail}
X.~Chen, N.~Mishra, M.~Rohaninejad, and P.~Abbeel, ``Pixelsnail: An improved
  autoregressive generative model,'' in \emph{International Conference on
  Machine Learning}.\hskip 1em plus 0.5em minus 0.4em\relax PMLR, 2018.

\bibitem{mesaros2016metrics}
A.~Mesaros, T.~Heittola, and T.~Virtanen, ``Metrics for polyphonic sound event
  detection,'' \emph{Applied Sciences}, 2016.

\bibitem{vadillo2022human}
J.~Vadillo and R.~Santana, ``On the human evaluation of universal audio
  adversarial perturbations,'' \emph{Computers \& Security}, 2022.

\bibitem{itu20031534}
B.~Series, ``Method for the subjective assessment of intermediate quality level
  of audio systems,'' \emph{International Telecommunication Union
  Radiocommunication Assembly}, 2014.

\bibitem{fad}
K.~Kilgour, M.~Zuluaga, D.~Roblek, and M.~Sharifi, ``Fr{\'e}chet audio
  distance: A reference-free metric for evaluating music enhancement
  algorithms.'' in \emph{INTERSPEECH}, 2019.

\bibitem{cramer2019look}
A.~L. Cramer, H.-H. Wu, J.~Salamon, and J.~P. Bello, ``Look, listen, and learn
  more: Design choices for deep audio embeddings,'' in \emph{ICASSP 2019-2019
  IEEE International Conference on Acoustics, Speech and Signal Processing
  (ICASSP)}.\hskip 1em plus 0.5em minus 0.4em\relax IEEE, 2019, pp. 3852--3856.

\bibitem{ChonGLI2023}
M.~Kang, S.~Oh, H.~Moon, K.~Lee, and B.~S. Chon, ``{FALL-E}: Gaudio foley
  synthesis system,'' Tech. Rep., June 2023.

\bibitem{GuanHEU2023}
S.~Fan, Q.~Zhu, F.~Xiao, H.~Lan, W.~Wang, and J.~Guan1, ``Foley sound synthesis
  with {A}udio{LDM} for dcase2023 task 7,'' Tech. Rep., June 2023.

\bibitem{Leemaum2023}
J.~Lee1, H.~Nam, and Y.-H. Park, ``{VIFS}: An end-to-end variational inference
  for foley sound synthesis,'' Tech. Rep., 2023.

\bibitem{ScheiblerLINE2023}
R.~Scheibler, T.~Hasumi, Y.~Fujita, T.~Komatsu, R.~Yamamoto, and K.~Tachibana,
  ``Class-conditioned latent diffusion model for {DCASE} 2023 foley sound
  synthesis challenge,'' Tech. Rep., 2023.

\bibitem{YiSURREY2023}
Y.~Yuan, H.~Liu, X.~Liu, X.~Kang, M.~D.Plumbley, and W.~Wang, ``Latent
  diffusion model based foley sound generation system for dcase challenge 2023
  task 7,'' June 2023.

\bibitem{BaiJLESS2023}
S.~Huang, J.~Bai, Y.~Jia, and J.~Chen, ``Jless submission to dcase2023 task7:
  Foley sound synthesis using non-autoagressive generative model,'' Tech. Rep.,
  June 2023.

\bibitem{ChangHYU2023}
W.-G. Choi and J.-H. Chang, ``{HYU} submission for the dcase 2023 task 7:
  Diffusion probabilistic model with adversarial training for foley sound
  synthesis,'' Tech. Rep., June 2023.

\bibitem{ChunChosun2023}
C.-W. Bang, N.~K. Kim, and C.~Chun, ``High-quality foley sound synthesis using
  monte carlo dropout,'' Tech. Rep., 2023.

\bibitem{ChungKAIST2023}
Y.~Chung, J.~Lee, and J.~Nam, ``Foley sound synthesis in waveform domain with
  diffusion model,'' Tech. Rep., 2023.

\bibitem{JungKT2023}
H.~C. Chung, Y.~Lee, and J.~H. Jung, ``Foley sound synthesis based on {GAN}
  using contrastive learning without label information,'' Tech. Rep., June
  2023.

\bibitem{KamathNUS2023}
P.~Kamath, T.~N. Islam, C.~Gupta, L.~Wyse, and S.~Nanayakkara, ``Dcase task-7:
  Style{GAN}2-based foley sound synthesis,'' Tech. Rep., June 2023.

\bibitem{LeeMARG2023}
K.~Kim, J.~Lee, H.~Kim, and K.~Lee, ``Conditional foley sound synthesis with
  limited data: Two-stage data augmentation approach with stylegan2-ada,''
  Tech. Rep., June 2023.

\bibitem{PillayCMU2023}
A.~Pillay, S.~Betko, A.~Liloia, H.~Chen, and A.~Shah, ``{DCASE} task 7: Foley
  sound synthesis,'' Tech. Rep., June 2023.

\bibitem{QianbinBIT2023}
A.~Qi, ``Auto-bit for {DCASE2023} task7 technical reports: Assemble system of
  bitdiffusion and {PixelSNAIL},'' 2023.

\bibitem{QianXuBIT2023}
H.~Zhang, K.~Qian, L.~Shen, L.~Li, K.~Xu, and B.~Hu, ``From noise to sound:
  Audio synthesis via diffusion models,'' 2023.

\bibitem{WendnerJKU2023}
T.~Wendner, P.~Hu, T.~Jadidi, and A.~Neuhauser, ``Audio diffusion for foley
  sound synthesis,'' Tech. Rep., June 2023.

\bibitem{XieSJTU2023}
Z.~Xie, X.~Xu, B.~Li, M.~Wu, and K.~Yu, ``The {X-LANCE} system for {DCASE2023}
  challenge task 7: Foley sound synthesis track b,'' Tech. Rep., June 2023.

\end{thebibliography}

\end{sloppy}
\end{document}